%% file: Formatting-Instructions-LaTeX-2022.tex
\def\year{2022}\relax
\documentclass[letterpaper]{article} % DO NOT CHANGE THIS
\usepackage{aaai22}  % DO NOT CHANGE THIS
\usepackage{times}  % DO NOT CHANGE THIS
\usepackage{helvet}  % DO NOT CHANGE THIS
\usepackage{courier}  % DO NOT CHANGE THIS
\usepackage[hyphens]{url}  % DO NOT CHANGE THIS
\usepackage{graphicx} % DO NOT CHANGE THIS
\usepackage{natbib}  % DO NOT CHANGE THIS AND DO NOT ADD ANY OPTIONS TO IT
\usepackage{caption} % DO NOT CHANGE THIS AND DO NOT ADD ANY OPTIONS TO IT
\DeclareCaptionStyle{ruled}{labelfont=normalfont,labelsep=colon,strut=off} % DO NOT CHANGE THIS
\usepackage{algorithm}
\usepackage{algorithmic}

\usepackage{newfloat}
\usepackage{listings}
\floatstyle{ruled}
\newfloat{listing}{tb}{lst}{}
\floatname{listing}{Listing}
\usepackage{appendix}
\usepackage{subfiles}
\usepackage{amsmath}
\usepackage{resizegather}
\usepackage{booktabs}
\usepackage{array}
\newcolumntype{C}[1]{>{\centering\arraybackslash}p{#1}}
\usepackage{subcaption}
\usepackage{xcolor}
\newcommand{\answerYes}[1]{\textcolor{blue}{#1}} 
\newcommand{\answerNo}[1]{\textcolor{teal}{#1}} 
\newcommand{\answerNA}[1]{\textcolor{gray}{#1}} 
 
\usepackage{placeins}

\title{Belief Alignment vs Opinion Leadership: Understanding Cross-linguistic Digital Activism in K-pop and BLM Communities}
\author{
    Yuheun Kim\textsuperscript{\rm 1},
    Joshua Introne\textsuperscript{\rm 1}
}
\affiliations{
    \textsuperscript{\rm 1}School of Information Studies, Syracuse University\\
    \{ykim72, jeintron\}@syr.edu
}

\begin{document}

\maketitle

\begin{abstract}
The internet has transformed activism, giving rise to more organic, diverse, and dynamic social movements that transcend geo-political boundaries. Despite extensive research on the role of social media and the internet in cross-cultural activism, the fundamental motivations driving these global movements remain poorly understood. This study examines two plausible explanations for cross-cultural activism: first, that it is driven by influential online opinion leaders, and second, that it results from individuals resonating with emergent sets of beliefs, values, and norms. We conduct a case study of the interaction between K-pop fans and the Black Lives Matter (BLM) movement on Twitter following the murder of George Floyd. Our findings provide strong evidence that belief alignment, where people resonate with common beliefs, is a primary driver of cross-cultural interactions in digital activism. We also demonstrate that while the actions of potential opinion leaders—in this case, K-pop entertainers—may amplify activism and lead to further expressions of love and admiration from fans, they do not appear to be a direct cause of activism. Finally, we report some initial evidence that the interaction between BLM and K-pop led to slight increases in their overall belief similarity.   
\end{abstract}

\section{Introduction}
\subfile{sections/1-Introduction}

\section{Background}
\subfile{sections/2-RelatedStudies}

\section{Methodology}
\subfile{sections/3-Methodology}

\section{Results}
\subfile{sections/4-Results}
\section{Discussion}
\subfile{sections/5-Discussions}

\section{Conclusion}
\subfile{sections/6-Conclusion}

\bibliography{aaai22}

\appendix
\section{APPENDIX}

\subsection{A. Probe Sentence Evaluation}
We developed six sentences that were compatible with our data but presented contextual nuances that might not be captured in existing language models. These were:
\begin{enumerate}
\item America has never fully addressed it's oppression for Black people.
\item There's no discrimination in America.
\item George Floyd died because of heart disease
\item George Floyd died because of police brutality
\item Bangtan is the best group.
\item Proof that Bangtan are beautiful
\end{enumerate}
We then examined the similarity scores for a subset of pairs in order to identify a model that was compatible with our intuitions. Scores for examined sentences pairs for each model testing are featured in Table \ref{tab:contextual}.
\begin{table*}[t]
\centering
% \resizebox{\linewidth}{!}{
\begin{tabular}{>{\raggedright\arraybackslash}p{0.5\linewidth}|cccccc}
\toprule
Model & (1) \& (2) & (3) \& (4) & (4) \& (5) & (5) \& (6) \\ \hline
symanto/sn-xlm-roberta-base-snli-mnli-anli-xnli \citep{mueller2022few} & 0.048      & 0.083  & -0.084  & 0.449   \\
sentence-transformers/LaBSE \citep{feng-etal-2022-language}      & 0.511      & 0.847    & -0.094  & 0.317 \\
togethercomputer/m2-bert-80M-32k-retrieval \citep{fu2023monarch}      & 0.170      & 0.483   & 0.333  & 0.314 \\
sentence-transformers/all-MiniLM-L6-v2 \citep{reimers-2019-sentence-bert}      & 0.527      & 0.745     & 0.052  & 0.561\\
sentence-transformers/all-mpnet-base-v2 \citep{reimers-2019-sentence-bert}      & 0.479      & 0.753    & 0.022  & 0.563\\
nomic-ai/nomic-embed-text-v1.5 \citep{nussbaum2024nomic}     & 0.673      & 0.839  & 0.352   & 0.746\\
hkunlp/instructor-xl \citep{su2023one}     & 0.676      & 0.820     & 0.516  & 0.785\\
intfloat/e5-large-v2  \citep{wang2022text}    & 0.790      & 0.914    & 0.73  & 0.846\\
thenlper/gte-large \citep{li2023towards}     & 0.841      & 0.907     & 0.7 & {0.907}\\
\bottomrule
\end{tabular}
% }
\caption{Similarity scores between probe sentences for each language model}
\label{tab:contextual}
\end{table*}

% \begin{table}[h]
% \centering
% % \resizebox{\columnwidth}{!}{
% \begin{tabular}{p{0.05\linewidth} | p{0.85\linewidth}}%{ll|c}
% \toprule
%  &   \textbf{Sentence}                               \\ \hline
% (a)& America has never fully addressed it's oppression for Black people.       \\
% (b)& There's no discrimination in America. \\
% (c)& George Floyd died because of heart disease                                                     \\
% (d)& George Floyd died because of police brutality                                                 \\
% (e)& Bangtan is the best group.                                                                     \\
% (f)& Proof that Bangtan are beautiful               \\                        
% \bottomrule
% \end{tabular}%
% % }
% \caption{Probe Sentences}
% \label{tab:probesentence}
% \end{table}

\subsection{B. Attractor Detail}
\begin{figure*}[!t]
\centering
\includegraphics[width=\textwidth]{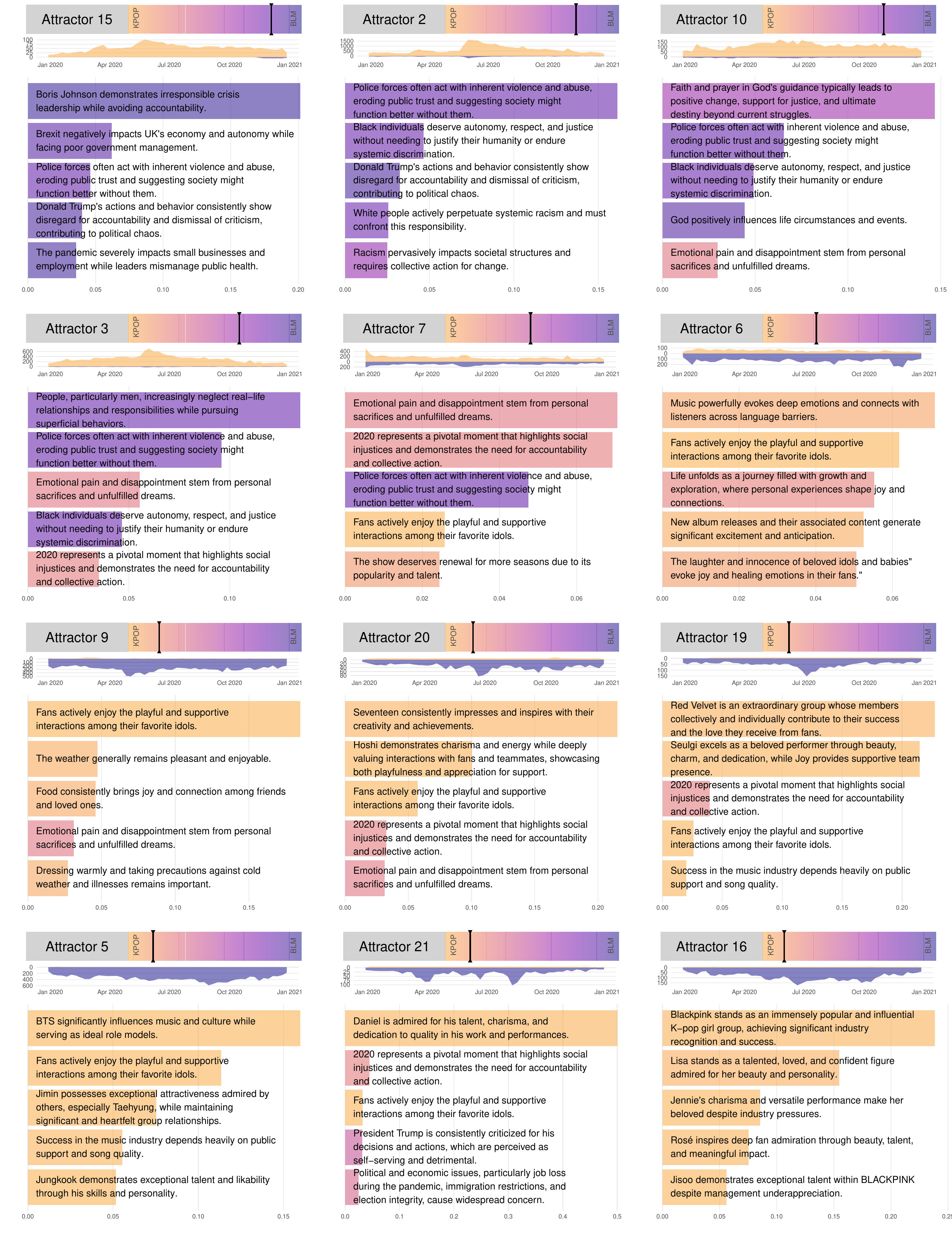} % Reduce the figure size so that it is slightly narrower than the column.
\caption{Visualizations of attractors mentioned in the main text. Only the top 5 belief summaries are shown for each attractor.}
\label{fig:attractor-dict}
\end{figure*}

Figure \ref{fig:attractor-dict} provides a visual summary of attractors that are referenced in the main body of the text. Attractors are sorted in reading order from BLM-biased to Kpop-biased. Each individual attractor visualization presents the following types of information, from top down:

\begin{enumerate}
    \item The overall bias of an attractor.
    \item Traffic fluctuations over time; the top portion indicates BLM, and the bottom portion indicates K-pop.
    \item The top five beliefs in the attractor; bars indicate relative intensity (the x-axis marks the relative overall intensity for each visualized belief) and colors indicate the bias of that belief, as described in the main body of the text.
\end{enumerate}

\subsection{C: Sensitivity Analysis}

\begin{figure}[ht]
    \centering
    \includegraphics[width=\columnwidth]{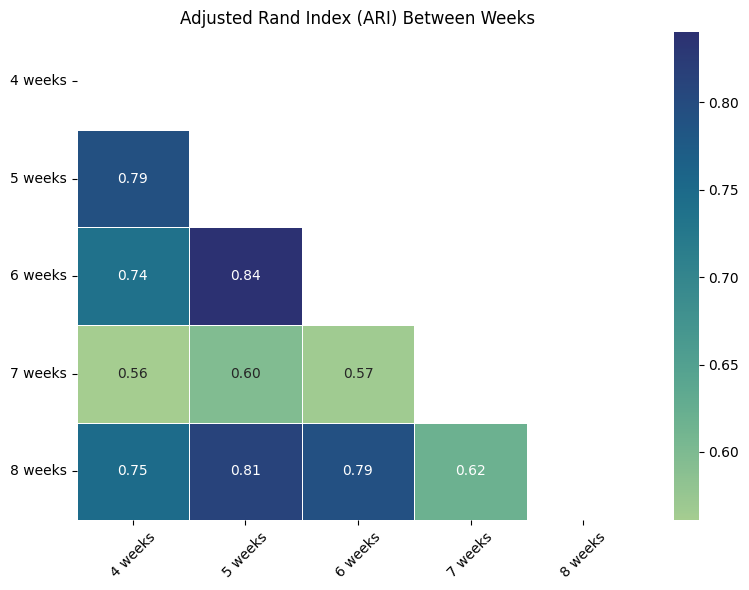}
    \caption{Heatmap of ARI scores between different half-life settings (weeks 4–8). Higher scores indicate greater similarity in attractor structure across time windows.}
    \label{fig:ARI-scores}
\end{figure}

We conducted a sensitivity analysis using a range of half-life values (4 to 8 weeks) to assess the stability of our results. For each setting, we re-generated belief vectors, identified density peaks, and reassigned attractor labels. Because attractor labels are arbitrary and non-aligned across settings, we used the Adjusted Rand Index (ARI) to evaluate the similarity of belief landscape structures. ARI, like Cohen’s Kappa, is a chance-adjusted measure of agreement, where values $\ge$ 0.6 indicate good alignment and values $\ge$ 0.8 reflect excellent agreement. Across the tested half-lives, ARI scores ranged from 0.56 to 0.84 (see Figure~\ref{fig:ARI-scores}), suggesting moderate to strong consistency in attractor structure.
To further test the robustness of our event-level findings, we examined whether the attractors that exhibited early spikes in the five-week setting—particularly attractors 2 and 3, which spiked following George Floyd’s murder but before BTS's tweet—had analogous counterparts in the other half-life configurations. Specifically, for each alternative half-life model, we computed the Jaccard similarity between attractors and those from the five-week model, identifying the most similar matches to attractors 2 and 3. We then checked whether these top-matching attractors also exhibited event spikes in the same temporal window.

Table~\ref{tab:jaccard-spikes} summarizes the Jaccard similarity scores and spike presence. In all cases, the attractor most similar to attractor 2 exhibited a spike. For attractor 3, its closest match exhibited a spike in two of the four models (weeks 4 and 7), but not in the others. No additional attractors showed spikes, reinforcing the specificity of the result.

These findings indicate that the core belief dynamics identified in our primary analysis—especially the concentration of early engagement in semantically mixed, BLM-leaning attractors—are robust across reasonable variations in the belief persistence parameter. While the precise attractor structure varies slightly with different half-life values, the key diffusion patterns remain consistent, strengthening confidence in our results.

\begin{table}[H]
\centering
\begin{tabular}{c|cccc}
\toprule
week5 & week4 & week6 & week7 & week8 \\
\midrule
attractor 2     & 0.74  & 0.86  & 0.55  & 0.81  \\
attractor 3     & 0.45  & -     & 0.82  & -   \\ 
\bottomrule
\end{tabular}
\caption{Top Jaccard similarity scores between attractors 2 and 3 from the 5-week setting and their best-matching attractors in alternative half-life models (weeks 4–8). Only attractors that also exhibited a spike in activity during the same time window are included.}

\label{tab:jaccard-spikes}
\end{table}

\subsection{D: Belief Cluster Validation}
We used the LLama 3.1 70B instruct model to evaluate the alignment of a sample of beliefs, sampled across all 266 identified beliefs, subject to the constraint that each assessed pair appeared in the same attractor. The prompt used was:

\begin{quote}
    The following two Tweets express beliefs about the world. I'd like you to determine whether or not these beliefs are well aligned. Each tweet may contain more than one belief, and if so, please only compare the beliefs that are most similar. Please code your comparison as follows:\\

    1 - The beliefs are mostly aligned with one another\\
    0 - None of the beliefs here can be readily compared\\
    -1 - The beliefs are mostly opposed with one another\\
\\
    Please format your response as a JSON string as follows:
    \begin{quote}
    \texttt{\{\\
    \ \ \ code: <your code>,\\
    \ \ \ explanation: <a brief explanation of your reasoning>\\
    \}}
    \end{quote}
\end{quote}
A sample of our results are included in Table \ref{tab:belief-pairs}.
\FloatBarrier
\begin{table*}[!t]
\centering
\begin{tabular}{p{0.27\textwidth} p{0.27\textwidth} c p{0.27\textwidth}}
\toprule
\textbf{Belief A} & \textbf{Belief B} & \textbf{Label} & \textbf{Explanation} \\
\midrule
President Trump has sent his congratulations to Attorney General William Barr after the government lawyers who prosecuted Trump's longtime confidant Roger Stone resigned in protest when their sentencing recommendation was slashed by the Justice Department. & Pelosi slams Trump for `abuse of power' in Roger Stone interventionBut the speaker declined to commit to a congressional investigation.  Remember, she waits until she has a case.  & -1 & The two tweets express opposing views on the same event. TWEET1 implies that President Trump's actions in the Roger Stone case are justified and worthy of congratulations, while TWEET2 portrays Trump's intervention as an `abuse of power' and criticizes it. \\
\addlinespace
Rebuttal: America needs a President that is responsible to human life. & Congress should immediately adopt provisions for remote voting so that they will be able to pass further bills in the face of this crisis. There are already 3 members of Congress who have tested positive and 31 self-quarantining and travel may become difficult. & 0 & The two tweets express unrelated beliefs about different aspects of governance. TWEET1 expresses a belief about the qualities required in a President, while TWEET2 proposes a specific policy solution for Congress. \\
\addlinespace
Fact Joe is creepy don’t vote Joe. Stick to the best president to ever take up residence in the White House. The only man that can save America is simply the best. And Donald despite what the British media says we love you over here. & Who needs more followers? Post your name below, follow every TRUMP supporter and RETWEET THIS! Follow me GO! Creepy Joe should be in jail! & 1 & Both tweets express strong support for Donald Trump and opposition to `Joe' (presumably Joe Biden). They use similar language, such as calling Joe `creepy', and both imply that Trump is the best option for America. \\
\bottomrule
\end{tabular}
\caption{Sample of Belief Pairs and Their AI-Annotated Alignment}
\label{tab:belief-pairs}
\end{table*}
\clearpage

\end{document}

%% file: sections/1-Introduction.tex
Digital technologies have reshaped the dynamics of social activism. In comparison to traditional social movements, \textit{digital activism} \citep{joyce2010digital} seems to bubble up organically from the fertile ground of stories shared and emotional connections made on social platforms \citep{papacharissi_affective_2015, bennett2003communicating}. However, the mechanisms by which digital activism spreads across cultural and linguistic boundaries remain unclear.

The events following George Floyd's murder by police on May 25, 2020, provide a unique window into these dynamics. On June 4, 2020, the Korean pop (K-pop) music group BTS and their management company tweeted their support for Black Lives Matter (BLM) while announcing a \$1 million donation to organizations associated with the movement. Within 24 hours, BTS's fandom---known as ARMY (Adorable Representative MC for Youth)---had mobilized through social media to match this donation. This rapid response raises a crucial question: Was the fandom's activism motivated by resonance with BLM's message, or were they simply following their idols' lead?

K-pop fandoms like ARMY represent a distinctly modern form of transnational community. Although rooted in South Korea, K-pop has become a global cultural phenomenon, with fans forming strong parasocial bonds with their favorite artists (known as ``idols'') \citep{yan2021parasocial}. These relationships are cultivated through highly interactive social media platforms, where idols maintain active presences and engage regularly with fans. ARMY, in particular, has attracted attention for its activism across a range of causes \citep{cho_bts_2022}, leading some to describe it as a ``leaderless movement'' \citep{carville_bts_2020}. Yet moments like ARMY’s swift matching of BTS’s donation to Black Lives Matter suggest that idols may still exert substantial influence over the political expression of their fan communities.

This case raises a broader question about how digital activism spreads across cultural and social boundaries. On one hand, digital platforms may foster organic diffusion by increasing the visibility of distant events and allowing communities to discover shared values. On the other, high-profile figures with large social media followings--such as K-pop idols--may act as opinion leaders, shaping how movements take root in new cultural contexts. Disentangling these mechanisms--organic value alignment versus opinion leadership--is key to understanding how global activism is coordinated in an increasingly connected world.

To explore this question, we examine social media discourse on Twitter / X from two distinct communities: Black Lives Matter activists and K-pop fans. We focus on the period surrounding George Floyd’s murder and BTS’s subsequent donation, applying the Belief Landscape Framework \citep{introne2023measuring}, a computational analysis of user-generated text data, to trace how beliefs evolved and diffused across these groups. Our analysis reveals that while both communities shared pre-existing concerns about police brutality, the data suggests that idol groups like BTS responded to, rather than initiated, these activist impulses. This implies that their influence may function less as directive leadership and more as amplification of already emerging sentiments.

This paper makes two primary contributions. The first is theoretical: we revisit long-standing questions about the cross-cultural diffusion of political beliefs in digital publics, using the convergence of BLM and K-pop activism as a unique case of dual engagement. The second is methodological: we introduce a novel application of the Belief Landscape Framework (BLF), which allows us to model belief dynamics across distinct social groups in a shared information environment. Although the focal event occurred in 2020, the analytical tools and core questions remain broadly applicable to other contexts of digital activism, cultural convergence, and online belief shifts.

In the following section, we situate our inquiry within theoretical and empirical literature on digital activism, social movements, and fan communities, developing specific hypotheses about how movements spread across cultural boundaries. We then present our methodology and findings, concluding with implications for understanding activism in an increasingly connected world.

%% file: sections/2-RelatedStudies.tex
Research on digitally enabled activism has documented how digital platforms enable movements to spread globally, transcending traditional geographic and linguistic boundaries. For example, \citet{bruns2013arab} found that Twitter's retweet functionality enabled information sharing across language communities during the Arab Spring, while \citet{chang2022justiceforgeorgefloyd} demonstrated how visual content on Instagram helped spread awareness of George Floyd's murder globally. However, while digital platforms clearly facilitate global spread, less is known about the proximal causes that drive cross-cultural interaction.

\subsection{Digital Activism and Social Movement Spread}
Digital media has fundamentally altered the dynamics of social movements by enabling new forms of participation, mobilization, and collective action \citep{tufekci_twitter_2017}. Bennett \& Segerberg \citeyearpar{bennett2012logic} argued that internet technologies allow individuals to engage through ``personalized action frames,'' which afford more diversity and autonomy in terms of engagement modes and the values, beliefs, and narratives that motivate action. Consequently, digital movements can be more inclusive, adaptable, and less reliant on traditional social movement organizations for coordination and curation.

Different theoretical frameworks suggest competing explanations for how these movements spread. Stern's \citeyearpar{stern_value-belief-norm_1999} value-belief-norm theory suggests that disparate groups become mobilized when they recognize shared values and beliefs expressed by a movement. The internet creates larger digital publics that span geopolitical boundaries, enabling individuals to connect with non-local events that resonate with their values \citep{castells_networks_2015}. These communities, forged around shared values, embody what \citet{anderson1991imagined} described as an ``imagined community.'' This sense of collective identity, as highlighted by \citet{melucci1996challenging} and \citet{gamson1992talking}, serves as a catalyst for collective action.

However, digital contexts also alter the power and dynamics of social influence. Preferential attachment in online networks produces individuals with enormous influence \citep{watts_influentials_2007}, who can serve as bridges between different cultural communities \citep{xu_predicting_2014}. These opinion leaders can motivate engagement \citep{roser2014genesis} and disseminate information to wider audiences \citep{chang2022justiceforgeorgefloyd}. In the context of global movements, local opinion leaders may serve as crucial bridges helping connect movements across geopolitical boundaries.

\subsection{Language and Culture in Digital Activism}
Digital platforms have created new forms of linguistic community that transcend traditional geographic boundaries. While geographic location once largely determined language use, online spaces enable the formation of language-based communities that maintain distinct cultural perspectives regardless of physical location \citep{lee_multilingual_2015,blommaert_durkheim_2018}. These communities often develop their own norms, practices, and shared cultural references, making language choice online a key marker of cultural identity and community membership.

K-pop fandom exemplifies these dynamics, with fans commonly using multiple languages to engage with content and each other \citep{alifa_multilingual_2024}. While Korean remains central to K-pop discourse, English serves as a bridge language, allowing fans to participate in global conversations while maintaining connections to Korean cultural contexts\citep{schneider_englishs_2024}. This multilingual practice creates unique opportunities to study how beliefs and values spread across linguistic communities.

\subsection{K-pop Fandom as Transcultural Community}
K-pop fandom represents a unique case study in how digital technologies reshape identity work and social influence in transnational contexts. Previous research has documented how K-pop fandoms develop distinctive cultural practices that transcend traditional geographic and national boundaries \citep{jin_social_2016, min_transcultural_2019}. This creates what Chin and Morimoto \citeyearpar{chin_towards_2013} call a ``transcultural fandom,'' where shared linguistic practices and cultural references create coherent communities spanning traditional cultural divides.

Recent work has examined ARMY's organizational dynamics during the \#MatchAMillion campaign specifically \citep{park_armed_2021}. Through survey research, Park et al. found that while ARMY lacks formal hierarchical structure, it maintains effectiveness through distributed ``pillar accounts'' that help coordinate action. Their findings suggest that shared values and shared identity, rather than direct leadership from BTS, were the primary motivator for participation in BLM support, though the fandom's established infrastructure enabled rapid mobilization.

\subsection{Research Questions}
Based on these competing theoretical frameworks and empirical findings, our first research question (\textbf{RQ1}) propose two hypotheses about how digital activism spreads across cultural boundaries:

\begin{quote}
\textbf{H1 (Belief alignment):} Online fans become activated to participate in a social movement because they resonate with beliefs that become salient in the movement.\\
\textbf{H2 (Opinion leadership):} Online fans become activated in response to explicit calls or behaviors by the influencers they follow.
\end{quote}

Additionally, we investigate whether cross-movement interaction creates lasting changes in expressed beliefs:

\begin{quote}
\textbf{RQ2 (Impact):} To what degree does cross-cultural interaction in the context of a pre-existing social movement create lasting changes in the expressed beliefs of the interacting groups?
\end{quote}

To investigate these questions, we employ the Belief Landscape Framework (BLF) developed by \citet{introne2023measuring}. The BLF provides advantages over previous approaches to studying belief dynamics in digital activism, which have relied primarily on qualitative discourse analysis \citep{bonilla_ferguson_2015}, hashtag tracking \citep{xiong_hashtag_2019}, or computational methods like topic modeling \citep{freelon2016beyond}. While these methods have yielded important insights, they are limited in their ability to trace how beliefs themselves spread and evolve. The BLF addresses these limitations by providing a quantitative approach specifically designed to track patterns of belief expression at scale while maintaining sensitivity to semantic content.

%% file: sections/3-Methodology.tex
We employed mixed methods to test our two hypotheses and investigate RQ2 (Impact), primarily using the Belief Landscape Framework (BLF) developed by \citet{introne2023measuring}. The BLF identifies ``belief attractors'' - clusters of users who express similar patterns of beliefs on Twitter. Attractors may be thought of a patterns of activation over a set of beliefs--a summary of the attractors used in this paper can be found in Appendix B.  These attractors act like gravitational wells in a belief space: users' expressed beliefs tend to become more similar to nearby attractors over time, and are statistically more likely to move toward them than away from them.

Each attractor is characterized by a set of commonly expressed beliefs and their relative intensities (based on expression frequency). While analogous to topics in topic modeling, attractors are composed of co-occurring belief statements rather than individual terms, providing a richer representation of shared belief patterns.

To evaluate H1 (Belief alignment), we first identified attractors that showed significant, coordinated spikes in activity from both K-pop and BLM communities following Floyd's murder but before BTS's tweet. We then analyzed the homogeneity of these attractors in the weeks preceding the murder. Finding that these activated attractors exhibited relatively low homogeneity (i.e., more mixed membership between communities) would support H1, suggesting that shared beliefs drove the initial response. We enriched this quantitative analysis with an AI-assisted summarization of the predominant beliefs within these attractors to understand which specific beliefs became activated.

For H2 (Opinion leadership), we examined whether BTS's tweet influenced K-pop fans to align more with BLM beliefs. This analysis faced two key challenges. First, because the tweet occurred only 10 days after Floyd's murder, high levels of activity could simply reflect the event's continued influence. Second, we cannot directly observe who saw the tweet or whether it influenced their behavior. To address these challenges, we conducted two analyses. Our primary analysis focused on members of the K-pop community who retweeted the original BTS tweet, as we can be certain these ``amplifiers'' engaged with it. 

Following established practices in computational social science \cite[e.g.,][]{barbera_birds_2015,darwish_unsupervised_2020}, we treat retweets as indicative of belief alignment or at least affiliative endorsement. While individual retweets may sometimes be sarcastic or critical, large-scale retweet patterns have been shown to accurately reflect political stance and social grouping. We therefore use retweet behavior as a proxy for belief-related engagement in our modeling framework, while acknowledging the inherent limitations of this assumption.

If opinion leadership drove belief change, we would expect to see these amplifiers move distinctly toward more BLM-aligned attractors after the tweet. Additionally, we examined whether the broader K-pop community showed significant spikes in activity around BLM-aligned attractors following the tweet, which could indicate viral spread of enthusiasm beyond the direct amplifiers. The procedure for establishing attractor alignment is detailed below.

Finally, to address RQ2 (Impact), we examined whether the murder and subsequent BTS tweet influenced the pattern of expressed beliefs for either community or their overall similarity with one another. Specifically, we delineate three periods of time: before the murder (``pre''), the period encompassing the murder and BTS tweet (``event''), and the time following these events (``post'') and calculated the mean tweet activity in each attractor for the two communities. We then examined Pearson correlations between periods within each community, as well as correlations between communities in each period.

The following subsections offer further detail on our data collection and analysis procedures.  All re-shareable data, code, and validation results can be found in a public GitHub repository accompanying this article\footnote{https://github.com/anonymized-for-blind-review}.

\subsection{Data Collection}
% Data Collection
We collected Twitter data using the Academic API (accessed May 2023) to analyze belief dynamics among K-pop fandoms and Black Lives Matter (BLM) activists across both short- and longer-term timeframes. The study centers on a pivotal moment of political convergence: the period following George Floyd’s murder, when both communities became visibly and vocally active. Notably, BTS and the ARMY fandom each made high-profile donations of one million dollars to BLM-related causes, signaling a surge of political engagement within the K-pop community.

Our sampling strategy was designed to capture this window of shared political expression. While K-pop fandoms and BLM activists differ markedly in their organizational structures and baseline levels of mobilization, this moment provided a rare opportunity to observe how beliefs might diffuse between two culturally distinct groups. Rather than assuming symmetry in ideology or structure, we focused on a common point of activation and sampled posts from users affiliated with each community. By tracing overlapping discourse themes during this period, we aim to understand how beliefs and narratives travel across social and cultural boundaries under conditions of heightened political attention.

Our data collection process began with gathering tweets from May-August 2020 containing either K-pop related hashtags (\#Kpop, \#KpopTwitter) or BLM related hashtags (\#BLM, \#BlackLivesMatter). From these tweets, we identified unique users and employed the Fasttext library \citep{joulin2016bag, joulin2016fasttext} to detect tweet languages. We retained only users tweeting predominantly in Korean, English, or both, while removing any self-identified bot accounts.

To create a balanced sample, we divided users into two groups based on whether their first post in our initial dataset appeared before or after George Floyd's murder on May 25, 2020. We then filtered for users who had between 500 and 50,000 tweets in 2020, ensuring sufficient data from each user while preventing any individual from dominating the sample. Through random sampling, we created a balanced dataset of 3,200 users with equal distribution between K-pop and BLM users, as well as between those who first posted before and after Floyd's murder.

For our final dataset, we retrieved all tweets from 2020 for each sampled user. After accounting for suspended and deleted accounts during this retrieval process, our final dataset comprised approximately 29 million tweets (15.5 million from BLM users and 13.6 million from K-pop users) from 2,948 users (1,460 BLM and 1,488 K-pop users). This balanced sampling approach enabled us to examine potential belief shifts surrounding George Floyd's murder while maintaining appropriate representation across both communities.

\subsection{Building the Belief Landscape}
% Data Processing
Our belief landscape methodology follows \citet{introne2023measuring}, with additional steps for Korean-to-English translation and dataset-specific adjustments.
\subsubsection{Translation}
The Belief Landscape Framework (BLF) from \citet{introne2023measuring} relies on an English-language dependency parser to extract belief statements (subject-verb-object tuples) and cannot process Korean text directly. We therefore translated all Korean tweets to English before applying the original parser.

We evaluated two translation models for this task. The MADLAD-400-3B model\footnote{https://huggingface.co/google/madlad400-3b-mt} \citep{kudugunta2024madlad}, Google's multilingual translation model, achieved a BLEU score of 30 despite the typical challenges of Korean-English translation pairs. We also tested KoBART \citep{skt-aikobart_2024}, which uses BART's encoder-decoder architecture but is trained on 40GB of Korean text. A Korean-English fine-tuned version of KoBART\footnote{https://github.com/seujung/KoBART-translation} achieved a higher BLEU score of 32.85.

The first author, a native Korean speaker, evaluated both models' performance on a sample of Korean tweets from our dataset. The KoBART model demonstrated more consistent and complete translations compared to MADLAD-400-3B, which often omitted portions of the source text. Based on this evaluation, we selected the KoBART model to translate all Korean tweets in our dataset.

\subsubsection{Identifying Prevalent Belief Subjects}
Following parsing, we processed the tweets to identify and consolidate predominant subjects from the subject-verb-object parse structures. This process involved removing empty and trivial subjects (such as pronouns and articles) and normalizing variants of the same subject (e.g., both `Donald Trump' and `Trump' were mapped to `Donald Trump'). We then identified the 100 most frequent subjects for each group (K-pop and BLM), yielding 183 unique belief subjects. For subsequent analysis, we retained only those tweets containing parsed beliefs about these subjects.

\subsubsection{Identifying Prevalent Belief Propositions}
Following \citet{introne2023measuring}, we mapped the retained tweets into a vector space using a language model. We evaluated nine open-source models trained on sentence similarity tasks, selecting candidates based on reported performance and relevance to our context. To identify the most suitable model, we tested each against a set of sentence pairs \citep{reimers-2019-sentence-bert}. Our evaluation focused on capturing intuitive semantic relationships - for example, ensuring that contradictory statements like ``America has never fully addressed its oppression of Black people'' and ``There's no discrimination in America'' were measured as semantically distant, while related statements like ``Bangtan is the best group'' and ``Proof that Bangtan are beautiful'' were measured as semantically similar. The complete evaluation results are presented in Appendix A Table \ref{tab:contextual}.

This analysis led us to select the XLM-RoBERTa-BASE model trained on Natural Language Inference (NLI) tasks \citep{mueller2022few}\footnote{https://huggingface.co/symanto/sn-xlm-roberta-base-snli-mnli-anli-xnli}. Using this model, we mapped all tweets into a vector space and reduced the dimensionality to two dimensions using UMAP \citep{mcinnes_umap_2020}. Our result revealed distinct clusters for K-pop and BLM users with areas of semantic overlap, indicating some shared beliefs between the groups. We then applied HDBScan clustering \citep{mcinnes_hdbscan_2017} to identify a total of 266 distinct belief clusters for subsequent analysis.

\subsection{Attractor Analysis}
To analyze belief dynamics, we constructed belief vectors following \citet{introne2023measuring}. One important parameter in this step of the algorithm is the choice of time window, which is used to account for the fact that people are likely to continue ``believing'' in a proposition for some time after it is uttered. To account for this, Introne \citeyear{introne2023measuring} applied an exponentially weighted moving average window to each user's tweet history, and examined results for consistency with theoretical predictions across a range of half-life values. Based on this analysis, Introne recommended using a exponential moving average with a five-week window.

\begin{figure}[ht]
\centering
\includegraphics[width=\columnwidth]{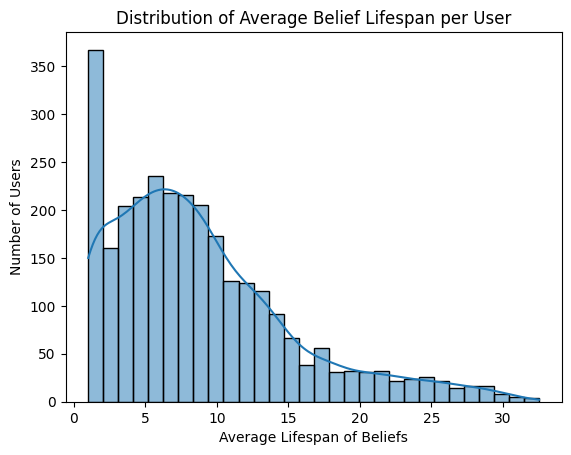} 
\caption{Distribution of average belief ``life-span''}
\label{fig:belief_lifespan}
\end{figure}

With Introne's results as context, we used a combination of empirical examination and sensitivity analysis to select a half-life for our analysis. First, we examined the distribution of observed belief ``lifespans'' in our data, where the lifespan of a belief is the span of time between the first and last mention of a given belief. Although the distribution is not normal (see Figure \ref{fig:belief_lifespan}), we observed a clear central tendency with a peak at roughly five weeks. We then examined a subset of our results (the structure of the belief landscape and event detection) across a range of half-lives from 4-8 weeks; details of this analysis are presented in Appendix C. Although there were some minor differences across this range of half-lives, support for our hypotheses is unchanged. Based on this, we adopt a five-week half-life for results reported here.

After establishing belief vectors, we embedded them in a two-dimensional space using UMAP \citep{mcinnes_umap_2020} and identified density peaks in the resulting distribution. This analysis revealed 21 distinct attractors.

\subsubsection{Validating Belief Coherence}

A known challenge with clustering beliefs in semantic embedding space is the risk of grouping semantically similar but stance-opposed statements (e.g., ``Climate change is a lie'' vs. ``Climate change is a threat'') \citep{introne2023measuring}. One way the Belief Landscape Framework (BLF) mitigates this is by evaluating belief vectors within the context of attractors—regions of behavioral convergence where individuals express belief patterns in common. This attractor-level structure offers a richer interpretive context, helping differentiate semantically similar beliefs that carry different stances.

To validate the coherence of belief clusters under this framework, we conducted a targeted evaluation of belief alignment among users co-located within the same attractor. From our full set of 266 belief categories—identified through prior clustering—we sampled up to 20 belief pairs per category, ensuring that both beliefs were expressed by users occupying the same attractor at the time. This yielded 5,294 belief pairs, proportionally distributed across attractors based on activity volume.

While we do not expect all clustered beliefs to be perfectly aligned in meaning, this sampling strategy offers a realistic test of whether beliefs grouped together in similar contexts reflect shared interpretations. Belief alignment was assessed using a combination of AI-based coding and manual validation. The LLaMA 3.1 70B model, run at a low temperature (0.2), was prompted to label each pair as aligned, incomparable, or opposed (the exact prompt used is presented in Appendix D). Manual review of a stratified subsample of 99 pairs yielded high inter-rater agreement (Cohen’s $\kappa = 0.82$), supporting the reliability of automated coding.

Overall, 71\% of belief pairs were coded as aligned, 16\% as incomparable, and 13\% as opposed; a sample of these results is provided in Appendix D. Many incomparable cases involved vague or emotionally expressive content with unclear belief structure (e.g., ``WAIT WHAT—OMG YUNHO CALM DOWN''), underscoring limitations in belief extraction for informal social text. Still, these results affirm that belief clusters, when grounded in attractor-level structure, reflect meaningful shared content rather than surface-level textual similarity alone. We reflect on these results more fully in the discussion section.

\subsubsection{Semantic Content Analysis}
To analyze the semantic content of each attractor, we developed a systematic labeling approach that combines automated text analysis with quantitative measures of user association. Each attractor represents a consistent pattern of belief expression, reflected in the relative frequencies with which users express semantically similar beliefs. To characterize each attractor's content, we first identified the ten tweets closest to the centroid of each belief cluster. We then used OpenAI's `gpt-4o-mini' language model to generate concise summaries of these representative tweets, providing a semantic interpretation of each belief cluster. Because attractors are derived from the same underlying set of 266 belief clusters, the same summarized beliefs appear in multiple attractors, but their relative frequencies vary.

\subsubsection{Relative Community Bias}
To support our analysis of H2, we developed a measure to quantify how predominantly each belief cluster was expressed by either the K-pop or BLM communities. For each belief cluster, we calculated a ``community bias score'' based on the relative frequency with which each community expressed such beliefs:
$$
bias = \frac{BLM_p}{BLM_{p} + KPOP_{p}}
$$
where $BLM_p$ and $KPOP_p$ represent the proportion of each community's total tweets that express beliefs in a given cluster. This normalization accounts for differences in overall activity levels between communities. A score of 1 indicates beliefs predominantly expressed within BLM discourse, while 0 indicates beliefs predominantly expressed within K-pop discourse.

Our use of tweet frequency rather than unique users is theoretically motivated. Communities' belief dispositions are manifested through their observable public discourse rather than individuals' internal states. Moreover, social media users actively curate their public expressions, aware of their role in collectively constructing their community's public face. Thus, the frequency of belief expression provides an appropriate measure of a belief's relative prominence within each community's discourse.

To characterize an entire attractor's bias, we computed the inner product between these bias scores and the normalized frequencies of beliefs within the attractor. This produces an overall measure of how predominantly an attractor's belief content is expressed within either community's discourse. Figure \ref{fig:attractor-dict} in Appendix B presents these bias scores along with the five most frequent beliefs for selected attractors.

\subsubsection{Weekly Attractor Homogeneity}
Distinct from the bias score, we also tracked the actual population of users occupying each attractor over time. At any given point, each user in our sample is associated with a single attractor based on their recent tweet history. While an attractor may be globally biased towards a given community, the actual mix of users expressing beliefs in that attractor space can vary significantly over time. The degree of mixing captures the dynamics of each community's week to week support for a given belief profile.

To evaluate H1, we quantified the degree of mixing using a homogeneity score, as follows. We labeled each attractor's weekly population mixture based on the relative tweet volumes from each community. Throughout our analysis, we bin data into weeks to control for systematic day-of-week variations in Twitter activity. For attractor $Attr_{n,w}$ in week $w$, let ${Kpop}_{n,w}$ and ${Blm}_{n,w}$ represent the number of unique active K-pop and BLM users respectively. The homogeneity score for an attractor in a given week is then:
\begin{equation}
\text{H}(Attr_{n,w}) = \left |\frac{Kpop_{n,w}-Blm_{n,w}}{Kpop_{n,w}+Blm_{n,w}}  \right |
\end{equation}

\begin{figure*}[ht]
\centering
\includegraphics[width=\textwidth]{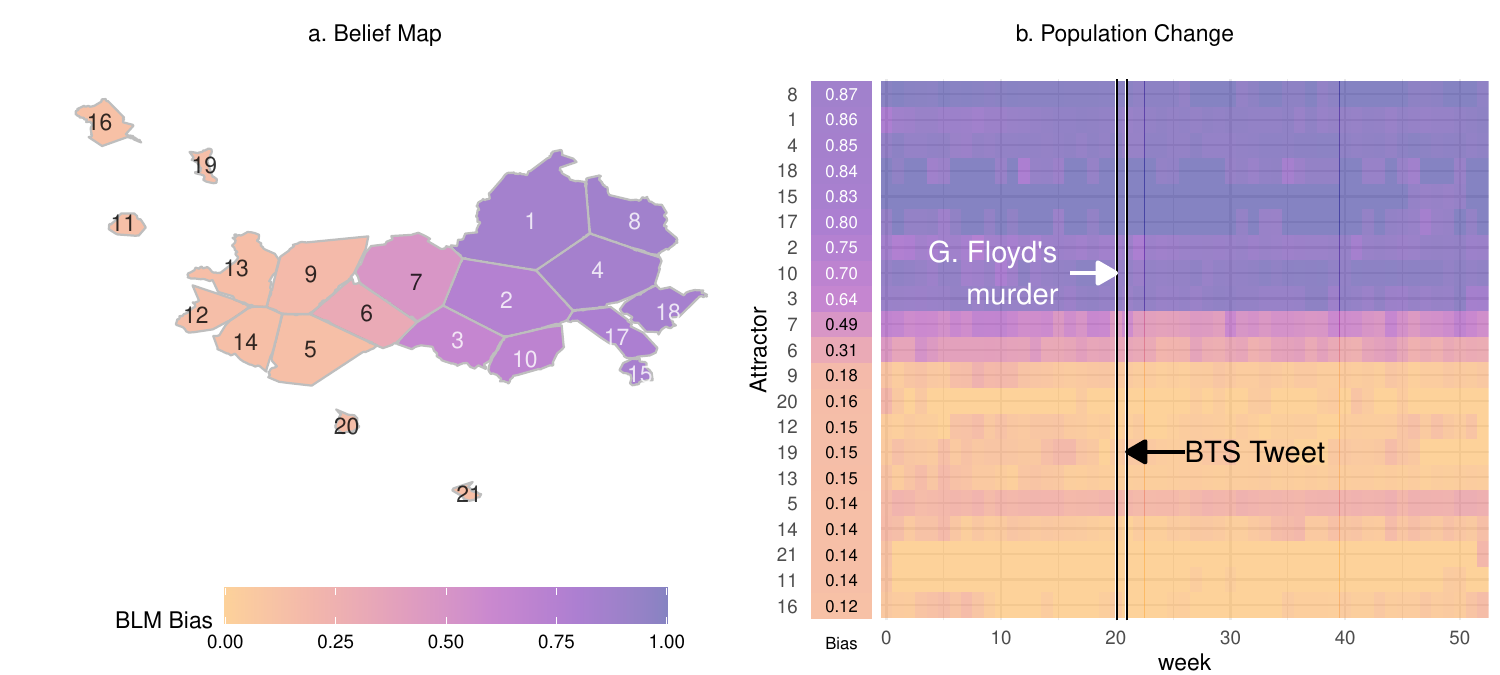} 
\caption{Summary information for the landscape and population dynamics. a) \textbf{Belief Map} - Spatial arrangement approximates semantic relationships among attractors, and color reflects the attractor bias score. b) \textbf{Population Change} -  Illustrates how the active population in each attractor changes from week to week. The y-axis corresponds to the individual attractors, sorted from bottom to top by bias score. Vertical lines mark weeks when the murder took place and BTS tweeted, in weeks 20 and 22, respectively.}
\label{fig1:overview}
\end{figure*}

Although not central for our analysis, we note that the distribution of homogeneity scores is heavily right-skewed (see Figure \ref{fig1:overview}), with only two attractors (6 and 7) having persistently low homogeneity scores (high degrees of mixing). We briefly discuss these two attractors in the Results section.

\subsubsection{Event Detection}
Event detection in social media streams, particularly Twitter, has been extensively studied \citep{li_event_2022}. Most approaches focus on the challenge of isolating semantically meaningful signals from high-volume, noisy data streams \citep[e.g,][]{weng_event_2011}. However, in our case, the Belief Landscape Framework has already performed this semantic filtering by identifying stable patterns of belief expression. Our task is therefore more focused: detecting meaningful spikes in population-specific activity within semantically coherent belief attractors.

This presents several methodological challenges. First, overall Twitter activity varies considerably on a weekly basis due to both seasonal patterns and exogenous events. Second, different belief attractors exhibit different baseline levels of activity. Third, the two populations we study (BLM and K-pop affiliated users) may have different characteristic patterns of activity. To address these challenges, we developed an approach that normalizes activity relative to both population-wide and attractor-specific baselines.

For each attractor $a$ and week $w$, we first calculate the proportion of each population's total activity occurring in that attractor:

\begin{equation}
    p_{a,w,pop} = \frac{x_{a,w,pop}}{\sum_i x_{i,w,pop}}
\end{equation}

where $x_{a,w,pop}$ is the observed activity (tweet count) for population $pop$ in attractor $a$ during week $w$, and the denominator sums over all attractors.

To establish baseline expectations while accounting for non-stationarity in the data, we use an exponentially weighted moving average (EWMA) with the same 5-week half-life used in constructing individual belief vectors. This alignment is important as it means our event detection operates at the same temporal scale as the belief dynamics we are studying. The expected proportion for week $w$ is:

\begin{equation}
    \hat{p}_{a,w,pop} = \alpha\sum_{i=1}^{w-1} (1-\alpha)^{i-1} p_{a,w-i,pop}
\end{equation}

where $\alpha = 1 - e^{\ln(0.5)/5}$ is derived from our chosen 5-week half-life.

The expected activity in an attractor for a given week is then:

\begin{equation}
    \hat{x}_{a,w,pop} = \hat{p}_{a,w,pop} \sum_i x_{i,w,pop}
\end{equation}

To identify significant deviations from these expectations, we calculate standardized scores using an exponentially weighted standard deviation:

\begin{equation}
    z_{a,w,pop} = \frac{x_{a,w,pop} - \hat{x}_{a,w,pop}}{\sigma_{a,w,pop}}
\end{equation}

where $\sigma_{a,w,pop}$ is the square root of the exponentially weighted variance:

\begin{equation}
    \sigma^2_{a,w,pop} = \alpha\sum_{i=1}^{w-1} (1-\alpha)^{i-1} (p_{a,w-i,pop} - \hat{p}_{a,w,pop})^2
\end{equation}

We identify significant spikes as those where $z_{a,w,pop} > 2$. This threshold corresponds roughly to a 95\% confidence interval under normal assumptions, though we note that the underlying distributions may be heavy-tailed. In practice, we find this threshold effectively identifies meaningful deviations while controlling for the varying baselines across attractors and populations.

%% file: sections/4-Results.tex
Figure \ref{fig1:overview} provides an overview of our attractor analysis. Figure \ref{fig1:overview}a visualizes the map along with attractor biases, while Figure \ref{fig1:overview}b captures both attractor biases and relative proportions of active users in each attractor over time. It is evident that Kpop and BLM attractors tend to be distinct and homogeneous, though several attractors are mixed and the mixing varies over time. AI-generated summaries for the attractors discussed below are available in Figure \ref{fig:attractor-dict}.

% \subsection{Quantitative Results}
\subsection{H1: Belief alignment}
\begin{figure*}[h]
\centering
\includegraphics[width=\textwidth]{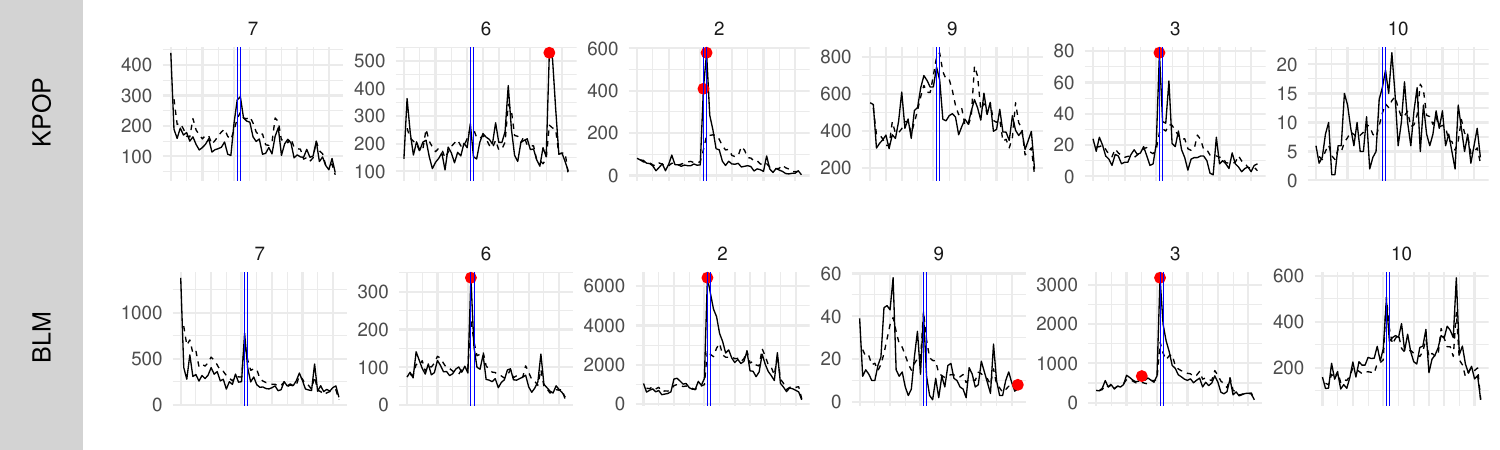} 
\caption{Summary of traffic and significant events for the six attractors exhibiting the most heterogeneity prior to the murder. The expected traffic is marked by a dashed line and significant events by dots. Vertical lines mark George Floyd's murder and the subsequent BTS tweet.}
\label{fig:weekly-users}
\end{figure*}
We sought support for H1 in mixed attractors that showed coordinated event spikes immediately following George Floyd's murder (week 20 in our dataset) but before the BTS tweet (week 22). Only two attractors (2 and 3) exhibited such coordinated spikes (see Figure \ref{fig:weekly-users}). Calculating the average homogeneity (H) of all attractors in the weeks prior to week 21, we find that these two attractors are among the most heterogeneous (Table \ref{tab:homogeneity}). 
\begin{table}[h]
\centering
\begin{tabular}{cc}
\toprule
attractor & homogeneity\\
\midrule
7 & 0.19\\
6 & 0.31\\
\textbf{2} & \textbf{0.74}\\
9 & 0.85\\
\textbf{3} & \textbf{0.87}\\
10 & 0.87\\
1 & 0.88\\
18 & 0.92\\
4 & 0.92\\
16 & 0.93\\
\bottomrule
\end{tabular}
\caption{Mean homogeneity scores for the top 10 most heterogeneous attractors prior to week 20.}
\label{tab:homogeneity}
\end{table}

\subsubsection{Qualitative inspection}
Summaries of the beliefs associated with attractors 2 and 3 are provided in Figure \ref{fig:attractor-dict}. Both focus on themes of race, gender, and police brutality, and are biased toward BLM-aligned beliefs. While K-pop users were underrepresented in these attractors prior to the murder of George Floyd—comprising just 7\% and 13\% of the active population in attractors 3 and 2, respectively—we observe a notable spike in their engagement with these regions during the week following the event.

Attractors 2 and 3 are not the most heterogeneous in absolute terms, but they are among the more mixed attractors in our distribution and are distinctly less polarized by group than the BLM-dominated attractors. Importantly, they occupy a semantically central region of the belief map, adjacent to other relatively heterogeneous attractors such as 6 and 7. Yet unlike those neighbors, attractors 2 and 3 lean toward BLM-related content and show meaningful cross-group engagement—especially from K-pop users during the event-driven window.

In contrast, attractors 6 and 7—while the most heterogeneous by homogeneity score (0.31 and 0.49, respectively)—are less clearly aligned with BLM discourse. Attractor 7 (42\% K-pop users in the week prior) includes beliefs about police brutality but also spans broader themes of personal hardship and social injustice, alongside K-pop content referencing idols and television. Attractor 6 (66\% K-pop users prior) is a high-entropy attractor defined by diffuse engagement across many beliefs, most prominently related to fandom, music, and celebrity culture. While these regions are mixed, they do not exhibit the same K-pop user surge during the critical period, nor are they clearly aligned with BLM frames.

This contrast reinforces our interpretation that belief alignment, rather than general activity or opinion leadership, best explains the observed diffusion. K-pop users did not engage randomly across mixed regions; instead, they were disproportionately drawn to attractors with BLM-aligned content that were not already saturated by either group—suggesting meaningful overlap in belief orientation rather than simple exposure effects.

In summary, we find moderate support for H1; although coordinated spikes do not appear in the most heterogeneous attractors, they do occur in attractors that are among the most heterogeneous in the weeks leading up to the murder, and these attractors highlight beliefs that are highly relevant to BLM and the overall context.

\subsection{H2: Opinion Leadership}
The BTS tweet addressing the BLM movement remained visible in attractors from week 22 through the end of the year, maintained by continued retweets from 272 users whom we refer to as ``amplifiers.'' To evaluate H2, we looked for significant shifts among these users toward BLM-oriented attractors following the BTS tweet.

\begin{figure}[h]
\centering
\includegraphics[width=\columnwidth]{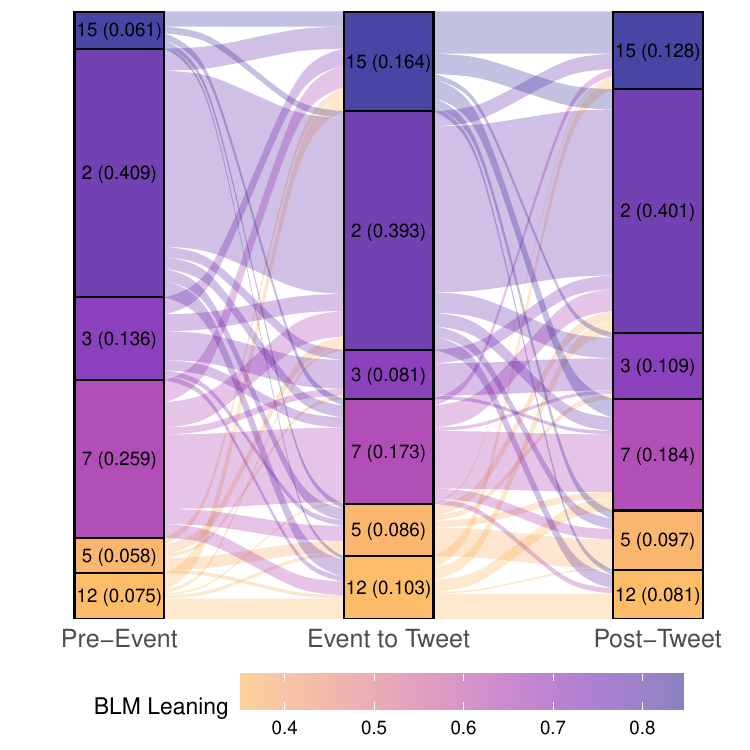} 
\caption{Amplifier flows across the belief landscape. The seven attractors shown account for approximately 90\% of all activity among these users in the weeks analyzed. Vertical strata are labeled with attractor id, and both strata size and parenthetical numbers indicate the proportional allocation of amplifiers. Color indicates the community bias of the attractor.}
\label{fig:flow-map}
\end{figure}

We tracked amplifier activity across six weeks: the two weeks before Floyd's murder (weeks 18 and 19), the two weeks including the murder but before the BTS tweet (weeks 20 and 21), and the two weeks following (weeks 22 and 23). While amplifiers were active in 18 of 21 attractors during this period, approximately 90\% of their activity concentrated in just 6 attractors. Figure \ref{fig:flow-map} visualizes these movements and each attractor's community bias.

Our analysis revealed little support for H2. While we observed a small increase in amplifier activity within BLM-biased attractors, these shifts occurred before the BTS tweet and subsequently diminished. The weighted average bias scores for attractors containing amplifiers across the three periods (pre-murder, between murder and tweet, and post-tweet) were 0.678, 0.68, and 0.674 respectively, indicating minimal movement. In fact, amplifiers appeared well-aligned with BLM-biased beliefs before the BTS tweet, tweeting these beliefs immediately after the murder but before the tweet itself.

We also examined event spikes within the broader K-pop community during the 10 weeks following the BTS tweet. Six attractors showed such spikes, but five were in highly K-pop biased attractors, focusing primarily on praise for individual performers and groups (attractors 20, 16, and 19) rather than social justice issues.

The exception was attractor 2, which spiked immediately following the earlier spike discussed in our H1 analysis. This sequence suggests that Floyd's murder initially activated K-pop community engagement with BLM issues, which BTS's tweet then amplified. We explore this interpretation further in the discussion section.

\begin{figure*}[h]
\centering
\includegraphics[width=\textwidth]{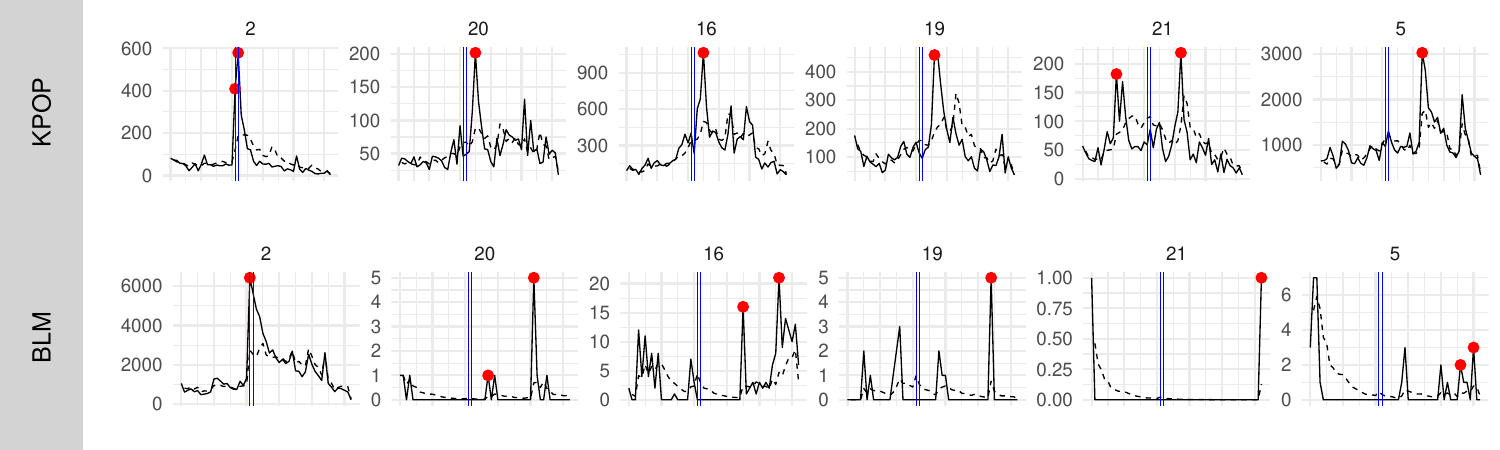} 
\caption{Summary of traffic and significant events for the six attractors with event spikes among K-pop users in the 10 weeks following the BTS tweet. The expected traffic is marked by a dashed line and significant events by dots. Vertical lines mark George Floyd's murder and the subsequent BTS tweet.}
\label{fig:weekly-users-h2}
\end{figure*}

\subsection{RQ2: Persistent Belief Change}
Results from our correlation analysis are shown in Table \ref{tab:correlations}. For both communities, the correlation patterns between periods show some variation but with overlapping confidence intervals, suggesting no statistically significant changes in how activity was distributed across attractors. For K-pop, while the event period correlation with post-period activity (r = 0.908) appears lower than pre-event to event correlation (r = 0.974), we cannot conclude this represents a meaningful shift. Similarly for BLM, despite apparent stronger correlation between event and post periods (r = 0.971) compared to pre-event and event (r = 0.904), these differences are not statistically significant.

Between community correlations were uniformly low (all negative) but this negative correlation weakened over time. There was a statistically significant difference (approximately p$<$.05) between pre-period (r = -0.171) and post-period traffic (r = -0.015). Based on this, there is a weak but meaningful difference in the direction of increasing similarity for the two communities. We return to this finding below.

\begin{table}[ht]
\centering
\begin{tabular}{llcc}
\hline
Group & Period & Correlation & 95\% CI \\
\hline
K-pop & Pre/Event & 0.974 & [0.936, 0.990] \\
 & Pre/Post & 0.940 & [0.855, 0.976] \\
 & Event/Post & 0.908 & [0.784, 0.963] \\
\hline
BLM & Pre/Event & 0.904 & [0.776, 0.961] \\
 & Pre/Post & 0.935 & [0.843, 0.974] \\
 & Event/Post & 0.971 & [0.929, 0.989] \\
\hline
Between & Pre & -0.171 & [-0.263, -0.077] \\
 & Event & -0.093 & [-0.333, 0.158] \\
 & Post & -0.015 & [-0.093, 0.064] \\
\hline
\end{tabular}
\caption{Correlation Analysis Results}
\label{tab:correlations}
\end{table}

%% file: sections/5-Discussions.tex
Our analysis provides novel insights into the drivers of cross-cultural digital activism, specifically in the context of online fandoms. Our findings support H1: belief alignment appears to have played a primary role in K-pop fandom's involvement in BLM. We found that K-pop fandom voiced shared beliefs about police brutality and racism with BLM before the event, while maintaining distinct dominant beliefs overall. This finding aligns with theoretical arguments suggesting that the digital transformation of activism enables more diverse, organic participation \citep{papacharissi_affective_2015,bennett2003communicating}. Cross-cultural digital activism appears driven by individuals' resonance with beliefs, values, and norms \cite{stern_value-belief-norm_1999} rather than opinion leadership.

Contrary to our initial hypothesis (H2), we find little evidence that BTS’s statement acted as a strong opinion leadership signal. Users who retweeted BTS exhibited only minor shifts in belief positions, much of which occurred prior to the tweet itself. This suggests that their engagement was likely shaped by broader discursive momentum rather than direct influence from BTS. These results underscore the importance of background informational context and suggest that visible leadership signals may not always function as causal drivers of belief change in networked publics.

While we found little support for H2, our analysis suggests opinion leadership operates in unexpected ways. The BTS tweet and management company's donation functioned more as an amplifying response to existing activism rather than an initiating event. The 272 ``amplifiers'' we identified may serve as hidden opinion leaders - these users demonstrated stronger alignment with BLM-biased beliefs than other K-pop users and potentially gained visibility to idols' management through high engagement. These results complement Park et al.'s \citeyearpar{park_armed_2021} findings about fandom structure. In particular, there may be significant overlap between our identified amplifiers and the ``pillar accounts'' they describe as coordinating fandom activities. This suggests a sophisticated dynamic where these accounts serve dual purposes: maintaining K-pop fandom's collective identity while simultaneously helping establish a social agenda that influences idol groups. Furthermore, our results indicate that when idols (and their management companies) respond to this agenda, it generates an outpouring of fan affection, creating a mutually reinforcing cycle that benefits both fans and artists.

Our investigation of RQ2 revealed two distinct but related findings. First, the belief structures within each community remained remarkably stable over time, with high correlations and overlapping confidence intervals indicating no substantial shifts in internal alignment. Second, we observed a small but statistically significant reduction in dissimilarity between the two groups' belief patterns, suggesting a subtle form of convergence. While intriguing, this convergence effect is difficult to interpret definitively: it may reflect a transient response to shared political events, or it could point to more enduring cross-group influence. Given the exploratory nature of this finding, we do not claim a causal relationship, but rather highlight the need for replication in other contexts where distinct cultural communities interact around shared political moments. While more sensitive statistical techniques could help uncover finer-grained dynamics, we believe the most important next step is to examine whether such convergence effects are robust across domains.

The Belief Landscape Framework enabled analysis impossible with previous approaches. While earlier studies focused on hashtags \citep{xiong_hashtag_2019} or content analysis \citep{raynauld_canada_2018}, the BLF allowed us to trace patterns of belief expression more directly. These methods offer valuable tools for studying social dynamics at scale. While our analysis centers on the specific case of K-pop and BLM in 2020, the methods we develop—particularly the use of belief landscapes to model intergroup diffusion—can be extended to a wide range of contexts. For example, they could be applied to understand how environmental narratives move across activist and industrial groups, or how health misinformation spreads between political ideologies. The generalizability lies in the structure of the approach: so long as belief-rich text data is available and group membership (or audience affinity) can be inferred, this framework can reveal how shared events reconfigure public discourse across divides.

Several limitations warrant consideration. As a single case study, generalizability remains uncertain. While our findings support our core hypotheses, alternative explanations remain plausible—for instance, increased K-pop fan engagement with BLM could reflect the broader sociopolitical climate of 2020, when pandemic lockdowns intensified online activity and racial justice awareness peaked globally.

Methodologically, our use of the Belief Landscape Framework (BLF) introduces both strengths and challenges. The framework requires several parameter choices, including the selection of a temporal window for belief extraction. We selected a five-week window based on prior work \citep{introne2023measuring} and our own analysis of belief lifespan distributions. While nearby windows yield comparable results, and five weeks represents a central tendency in our data, the lack of a principled theory for optimal window size remains a limitation—particularly given likely variation across individuals and beliefs.

We also undertook a validation of belief coherence to assess the alignment of clustered beliefs. While our results suggest that belief clusters are reasonably coherent when filtered through attractor structure, the analysis revealed some limitations in the belief extraction process. A nontrivial portion of tweets lacked clear belief content, and semantic clustering occasionally grouped beliefs with divergent stances. The use of co-occurrence structure helps mitigate these issues, but noise remains. We believe that future applications could benefit from more powerful AI-assisted belief extraction to improve cluster resolution.

Finally, information loss due to Korean-English translation and the increasingly restricted nature of platform APIs complicate reproducibility. While \citet{introne2023measuring} found the BLF to be robust across parameter ranges, the method remains novel, and best practices are still emerging. These challenges underscore the need for broader community guidelines around parameter selection, belief extraction, and historical dataset sharing in social media research.

Finally, ethical considerations extend beyond basic privacy protections. While we've avoided reproducing individual tweets (with the exception of the Appendix, wherein we thought it important to include raw tweets in order to illustrate some of the nuances of belief therein), the large-scale analysis of beliefs raises broader concerns about potential misuse. We advocate thoughtful engagement with these methods while acknowledging their value for understanding social movements.

%% file: sections/6-Conclusion.tex
Our study suggests that belief alignment, rather than opinion leadership, primarily motivates cross-cultural digital activism in online fandoms. Through the Belief Landscape Framework analysis of K-pop fans' interaction with the Black Lives Matter movement on Twitter, we found that digital transformation enables diverse and organic participation, with individuals resonating with emergent beliefs and values that transcend geo-political boundaries.

Our analysis reveals nuanced dynamics between online fandoms and their idols. While K-pop entertainers' actions catalyzed expressions of support from fans, they amplified rather than initiated activism. The pre-existing values and beliefs of fans themselves drove engagement with social causes.

The BLF proved instrumental in identifying local patterns of belief formation, offering advantages over traditional hashtag or content analysis. Despite challenges with cross-linguistic analysis and methodology refinement needs, this approach effectively illuminates the complex interplay of beliefs, values, and norms in online communities. The methodological approach we introduce is not limited to the K-pop/BLM case, but provides a generalizable toolkit for understanding how beliefs evolve and diffuse across social boundaries in response to politically salient events. As such, it offers a foundation for future work exploring the dynamics of cultural convergence, political polarization, and networked collective behavior in diverse digital communities.

Looking ahead, the BLF and related methods hold promise for studying socio-cultural opinion dynamics. These approaches enable researchers to track the emergence, evolution, and interaction of beliefs across diverse online communities, providing insights into social change and cross-cultural engagement in the digital age. As digital platforms increasingly shape public discourse and collective action, we believe that tools for mapping and analyzing belief landscapes will become valuable for understanding social movements, cultural exchange, and political mobilization.